\documentclass[aps,prl,article,twocolumn,citeautoscript,footinbib,superscriptaddress]{revtex4-2} \synctex=1 
\usepackage{amsmath,amssymb,bm,bbm} 
\usepackage{amsthm}
\usepackage{physics}
\usepackage{graphicx}  
\usepackage{color} 
\usepackage[dvipsnames]{xcolor}
\usepackage[colorlinks=true]{hyperref}  
\usepackage[section]{placeins}
\usepackage[mathscr]{euscript}
\usepackage[toc,page]{appendix}
\hypersetup{
    unicode=false,          
    pdftoolbar=true,        
    pdfmenubar=true,        
    pdffitwindow=false,     
    pdfstartview={FitH},    
    pdftitle={},    
    pdfauthor={},     
    pdfsubject={},   
    pdfcreator={},   
    pdfproducer={}, 
    pdfkeywords={} {} {}, 
    pdfnewwindow=true,      
    colorlinks=true,       
    linkcolor=magenta, 
    citecolor=blue,        
    filecolor=magenta,      
    urlcolor=blue           
} 

\usepackage{pdfpages}
\usepackage{pgffor}
\makeatletter
\AtBeginDocument{\let\LS@rot\@undefined}
\makeatother

\usepackage{tikz}
\usepackage{tikz-cd}
\usetikzlibrary{arrows}
\usetikzlibrary{intersections}
\usetikzlibrary{shapes.geometric}
\usetikzlibrary{decorations.pathmorphing, patterns,shapes}

\renewcommand{\bar}{\overline}
\renewcommand{\tilde}{\widetilde}

\renewcommand{\leq}{\leqslant}

\newcommand{\SU}{\operatorname{SU}}

\usepackage{amsfonts}
\usepackage{upgreek}
\usepackage{slashed}
\usepackage{latexsym}
\usepackage{changes}

\newcommand{\beq}{\begin{equation}}
\newcommand{\eeq}{\end{equation}}
\def\bea{\begin{eqnarray}}
\def\eea{\end{eqnarray}}
\newcommand{\sgn}{\operatorname{sgn}}

\begin{document}
\title{Quantum phase transition at non-zero doping in a random $t$-$J$ model}

\author{Henry Shackleton}
\affiliation{Department of Physics, Harvard University, Cambridge MA 02138, USA}

\author{Alexander Wietek}
\affiliation{
Center for Computational Quantum Physics, Flatiron Institute, New York, NY 10010 USA
}

\author{Antoine Georges}
\affiliation{
Center for Computational Quantum Physics, Flatiron Institute, New York, NY 10010 USA
}
\affiliation{Coll{\`e}ge de France, 11 place Marcelin Berthelot, 75005 Paris, France}
\affiliation{CPHT, CNRS, {\'E}cole Polytechnique, IP Paris, F-91128 Palaiseau, France}
\affiliation{DQMP, Universit{\'e} de Gen{\`e}ve, 24 quai Ernest Ansermet, CH-1211 Gen{\`e}ve, Suisse}
\author{Subir Sachdev}
\affiliation{Department of Physics, Harvard University, Cambridge MA 02138, USA}

\begin{abstract}
We present exact diagonalization results on finite clusters of a $t$-$J$ model of spin-1/2 electrons 
with random all-to-all hopping and exchange interactions.
We argue that such random models capture qualitatively the strong local correlations needed to describe the cuprates and related compounds, while avoiding lattice space group symmetry breaking orders. The previously known spin glass ordered phase in the insulator at doping $p=0$ extends to a metallic spin glass phase up to a transition $p=p_c \approx 1/3$.
The dynamic spin susceptibility  shows signatures of the spectrum of the Sachdev-Ye-Kitaev models near $p_c$. We also find signs of the phase transition in the entropy, entanglement entropy and compressibility, all of which exhibit a maximum near $p_c$. The electron energy distribution function in the metallic phase is consistent with a disordered extension of the Luttinger-volume Fermi surface for $p>p_c$, while this breaks down for $p<p_c$.
\end{abstract}
\maketitle
Two recent experiments \cite{Ramshaw20,Julien19} have shed new light on the transformation in the metallic parent state of the cuprate superconductors near optimal doping, while also highlighting the central theoretical puzzles. Angle-dependent magnetoresistance measurements in La$_{1.6-x}$Nd$_{0.4}$Sr$_x$CuO$_4$ \cite{Ramshaw20} are compatible with a Luttinger volume `large' Fermi surface only at a hole doping $p>p_c \approx 0.23$. Nuclear magnetic resonance and sound velocity measurements in La$_{2-x}$Sr$_x$CuO$_4$ \cite{Julien19} in high magnetic fields have uncovered glassy antiferromagnetic order for $p < p_c \approx 0.19$. These, and other, observations show that the parent metallic state of the cuprates exhibits Fermi liquid behavior for $p>p_c$, and transforms to an enigmatic pseudogap metal with glassy magnetic order for $p<p_c$. Observations also indicate that the reshaping of the Fermi surface, and the onset of the pseudogap, for $p<p_c$ cannot be explained 
by long-range antiferromagnetic order, which sets in at a doping smaller than $p_c$.

Here, we present exact diagonalization results on clusters of $N$ sites of a $t$-$J$ model with random and all-to-all 
hopping and exchange interactions (see (\ref{eq:randomtJ})). In the thermodynamic limit $N\rightarrow \infty$, the replica-diagonal saddle point of this model, and a related Hubbard model \cite{Cha19}, are described by (extended) dynamic mean-field equations in which the disorder self-averages~\cite{supp}. 
Moreover, closely related mean-field equations also appear in non-random models in high spatial dimensions \cite{Smith2000,Haule02}, indicating that the self-averaging features of the random models properly capture generic aspects of strong correlation physics. A direct solution of the $N=\infty$ replica-diagonal saddle point of the Hubbard model is presented in a separate paper \cite{dumi2020}, with complementary results which are consistent with our conclusions below.

The insulating model at $p=0$ has been studied previously by exact diagonalization \cite{MJR02}, 
and a non-self-averaging spin glass ground state was found.
We find similar results at $p=0$, but with a reduced estimate for the magnitude of the spin glass Edwards-Anderson order parameter, $q$. At non-zero $p$, we find that $q$ decreases monotonically, vanishing at a quantum phase transition $p_c \approx 1/3$. We present several results for thermodynamic, entanglement, and spectral properties across this transition. All our results are consistent with the presence of a self-averaging Fermi liquid state for $p>p_c$; in particular, we find that the one-particle energy distribution function is consistent with a disordered analog of the Luttinger theorem~\cite{supp}. The entropy, entanglement entropy 
and compressibility all have maxima near $p_c$. We find that the low frequency dynamic spin susceptibility matches that of the Sachdev-Ye-Kitaev (SYK) class of models \cite{SY92,kitaev2015talk}
over a significant range of frequencies near $p_c$; this includes a subleading contribution which arises from a boundary graviton in dual models of two-dimensional quantum gravity \cite{SS10,Maldacena:2016hyu,Kitaev:2017awl,MariaI20,TGSTtJ}. Such spectral features are not present in theories that treat the transition at $p=p_c$ in a Landau-Ginzburg-Hertz framework for the onset of spin glass order in a Fermi liquid \cite{Sengupta95,SRO1995}. 

{\it Random $t$-$J$ model.} 
We consider the Hamiltonian
\begin{equation}
    H = \frac{1}{\sqrt{N}} \sum_{i \neq j = 1}^N t_{ij} P c^\dagger_{i \alpha} c_{j \alpha} P + \frac{1}{\sqrt{N}} \sum_{i < j = 1}^N J_{ij} \vb{S}_i \cdot \vb{S}_j
    \label{eq:randomtJ}
\end{equation}
where $P$ is the projection on non-doubly occupied sites, $\vb{S}_i = (1/2) c_{i \alpha}^\dagger \vb{\sigma}_{\alpha \beta} c_{i \beta}$ is the spin operator on site $i$. The hoppings $t_{ij} = t_{ji}^*$ and real exchange interactions $J_{ij}$ are independent random numbers with zero mean and variance $t^2, J^2$. Henceforth, we set $t=J=1$. We work in the canonical ensemble, where our system has a fixed particle (hole) density, $n$ ($p = 1-n$).
At $p=0$, hopping is prevented due to the double occupancy constraint, and the model reduces to an infinite-range Heisenberg model with random couplings. The $p=0$ model has been studied analytically by generalizing the $\SU(2)$ symmetry to $\SU(M)$ and taking a large-$M$ limit~\cite{SY92, GPS00, GPS01}, and numerically for the case of $M=2$ ~\cite{MJR02, MJR03}. For $\SU(2)$, a spin glass phase is found below a critical temperature $T_c \approx 0.10J$. When doping is present, Ref.~\cite{PG98} predicts a disordered Fermi liquid phase for all non-zero values of $p$ in the large-$M$ limit. However, it was recently argued~\cite{randomtJ1, randomtJ2} that for the case of $\SU(2)$, the spin glass phase should persist up to a critical finite value of doping, $p_c$, corresponding to a quantum critical point separating the spin glass phase from a disordered Fermi liquid.
Near criticality, the model is predicted to exhibit SYK-like criticality with a non-zero extensive entropy and a linear-in-temperature resistivity~\cite{Guo2020}. In a weak-coupling renormalization group, this critical point emerges when the three fractionalized excitations in the $t$-$J$ model become degenerate in energy, leading to a zeroth order prediction of $p_c = 1/3$. 

{\it Dynamical Spin Response at $T=0$.} 
We first present results on the nature of the spin correlations at $T=0$, providing evidence that the spin glass phase shown to exist at $p=0$ is stable for small values of doping, up to a critical value of doping near $p=1/3$. Using the Lanczos algorithm, we calculate the spectral function at $T=0$, 
\begin{equation}
\begin{aligned}
    \chi''(\omega) &= \frac{1}{3} \sum_\alpha \frac{1}{N} \sum_i \sum_n \abs{\bra{\psi_n} S_i^\alpha \ket{\psi_0}}^2 
    \\
    &\times \left[\delta(\omega - (E_n - E_0)) - \delta(\omega + (E_n - E_0))\right]\,,
    \label{eq:spectralFunction}
    \end{aligned}
\end{equation}
where numerically the delta functions are replaced by Gaussians with a small variance. 
The signature of spin glass order,
$\lim_{t \rightarrow \infty} \frac{1}{N}\sum_i \langle \vb{S}_i(0) \vb{S}_i(t) \rangle = q \neq 0$, is reflected by a $q \delta(\omega)$ contribution to the dynamical structure factor $S(\omega)$, which is related to the spectral function at $T=0$ by $\chi''(\omega) = S(\omega) - S(-\omega)$. 
For a finite system size, the exact delta function in $S(\omega)$ is replaced by a peak at low frequency, whose width approaches $0$ in the thermodynamic limit and whose total spectral weight gives $q$. Therefore, the spin glass contribution to $\chi''(\omega)$ for finite systems is given by a low frequency peak, and was analyzed for this model at $p=0$ in~\cite{MJR02}. Above $p_c$, a disordered Fermi liquid is expected to have a low-frequency behavior of $\chi''(\omega) \sim \omega$. 

\begin{figure}
    \centering
    \includegraphics[width=0.45\textwidth]{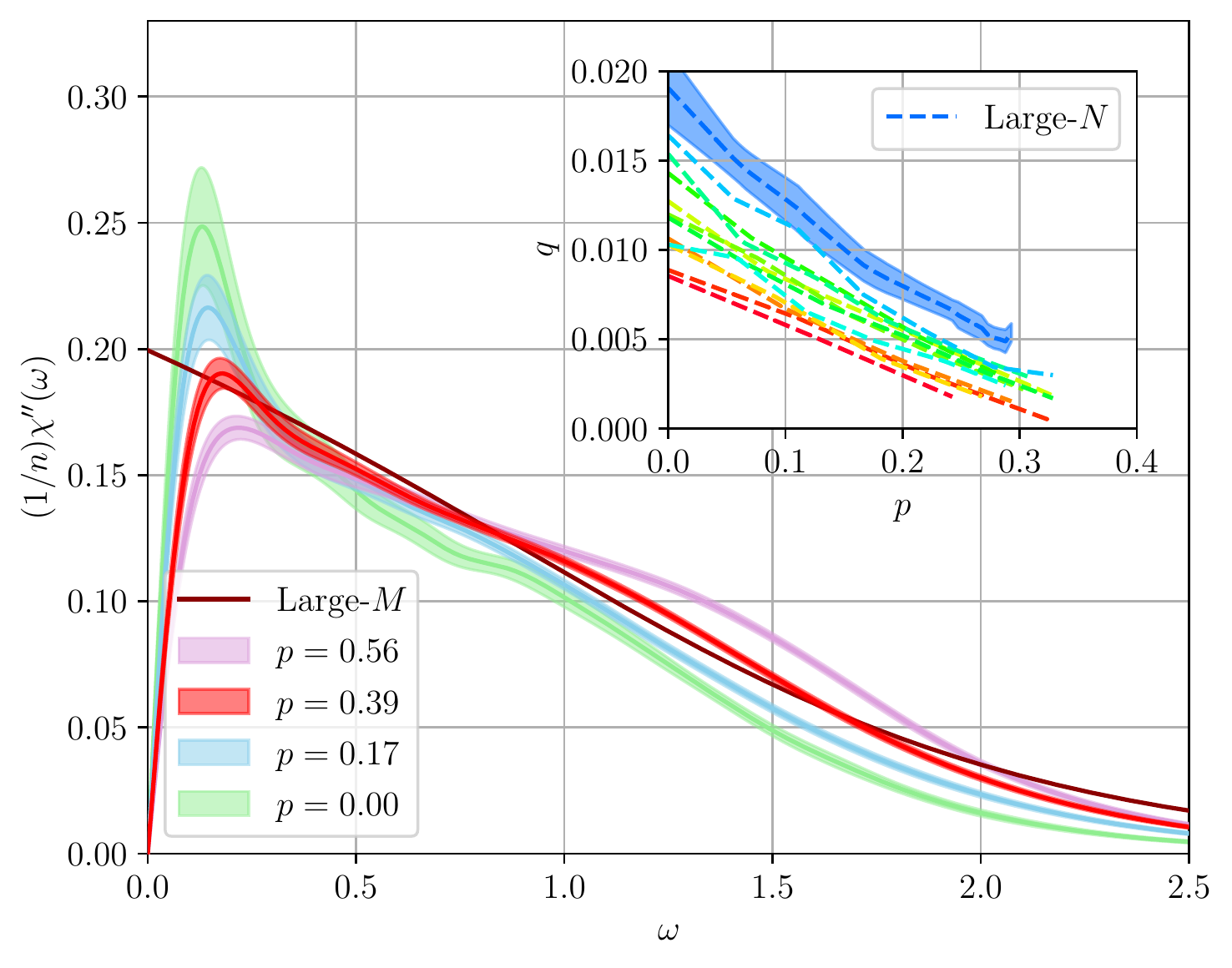}
    \caption{
    The spectral function $\chi''(\omega)$ of the random $t-J$ model, averaged over 100 disorder realizations on an 18-site cluster. At low dopings, a sharp peak at low-frequency at low doping is indicative of spin glass order. With increasing doping, the magnitude of this peak is reduced, and the low-frequency behavior closely resembles the rescaled spectral function of  the large $M$ SYK theory \cite{SY92,MariaI20,TGSTtJ}. (Inset) After an extrapolation to the thermodynamic limit, the integrated weight of the low-frequency peak is non-zero, indicating spin glass order. This weight vanishes near $p\approx 0.4$. Plotted is the integrated weight for $8 \leq N \leq 18$ (as a gradient from red to blue), and the large-$N$ extrapolation with error bars.}
    \label{fig:spectralFunction}
\end{figure}

\begin{figure*}[t]
    \centering
    \includegraphics[width=\textwidth]{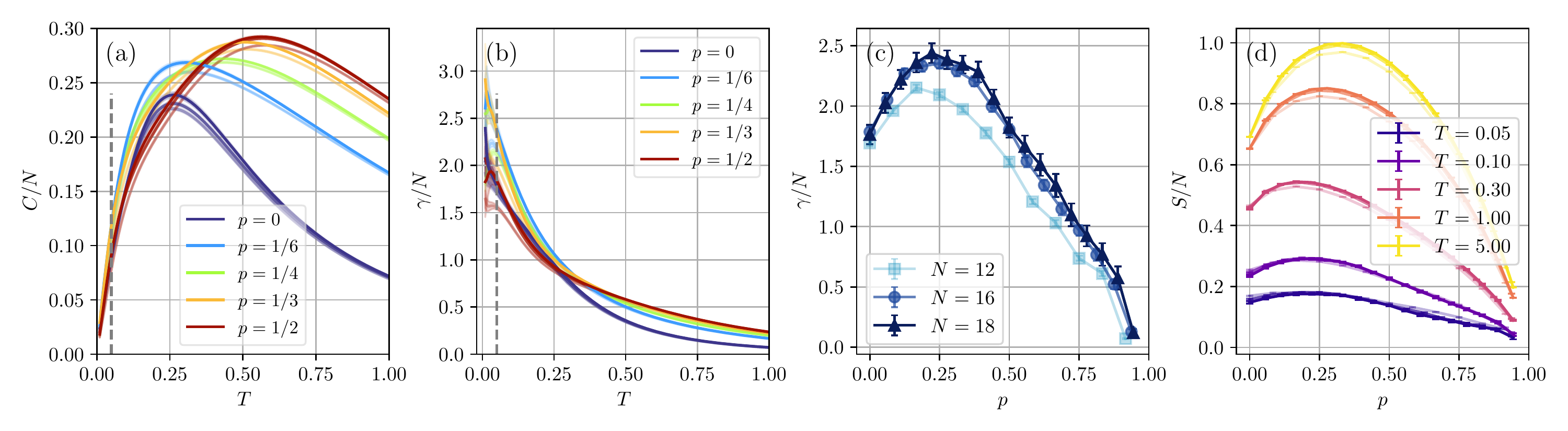}
    \caption{Thermodynamics of the random $t$-$J$ model for system sizes $N=12, 16, 18$, indicated by increasing opacity.
    (a) The specific heat $C$ as a function of temperature for various values of doping. 
    (b)  The linear-in-$T$ coefficient of specific heat, $\gamma = C/T$, for various
    dopings as a function of temperature, and (c) for $T=0.05$ as a function of
    doping. 
    (d) The thermal entropy $S$ as a function of doping for various temperatures.}
    \label{fig:thermodynamics}
\end{figure*}

The spectral function for the random $t$-$J$ model, calculated using the Lanczos algorithm on an 18-site cluster, is shown for several values of doping in Fig.~\ref{fig:spectralFunction}. A prominent hump at low-frequency for dopings $p \lesssim 0.4$ suggests the presence of spin glass order in this range of doping. However, a large-$N$ analysis of this hump must be performed in order to verify that the hump asymptotes to a delta function in the thermodynamic limit. To do this, we first subtract off a background contribution to account for the rest of the spectral weight. Anticipating SYK behavior near the critical point at low frequencies, we subtract a spectral weight obtained by rescaling the solution of the Schwinger-Dyson equations of the $p=0$ model in the large-$M$ limit \cite{SY92,MariaI20,TGSTtJ} (we rescale $J$, while preserving total spectral weight). This SYK spectral weight has a leading term $\chi''(\omega) \sim \sgn(\omega)$ as $|\omega| \rightarrow 0$ at $T=0$ (which generalizes to $\tanh\left({\omega}/{2T}\right)$ at low $T$). 
The next-to-leading SYK term depends linearly in $\omega$, and arises
from the boundary graviton in the holographic dual~\cite{MariaI20,TGSTtJ}. 
It is important to note that the exponents of these two leading SYK contributions are universal and independent of $M$.
Away from the critical point and in the spin glass phase, we find that the spectral function is described well by a combination of the SYK result and a low-frequency hump. A large-$N$ analysis of this low-frequency hump, described in more detail in the supplementary material, confirms that the variance of the hump vanishes in the thermodynamic limit, whereas the spectral weight, shown in Fig.~\ref{fig:spectralFunction}, remains non-zero. Our analysis gives a large-$N$ estimate of $q \sim 0.02$ at $p=0$. For larger values of doping, $q$ decreases from its value at $p=0$, eventually vanishing at some critical value of doping $p_c$. By linearly extrapolating the large-$N$ prediction for $q$ to higher dopings, we obtain an estimate of $p_c = 0.420 \pm 0.007$. Around this range of dopings, the spectral function shows good agreement with the large-$M$ critical prediction given in Fig.~\ref{fig:spectralFunction}. At dopings well above $p=0.4$, we find the spectral function to be largely independent of system size. No gap at low frequency is visible, and $\chi''(\omega) \sim \omega$ behavior consistent with Fermi liquid predictions is clear. We will provide a more rigorous verification of the Fermi liquid phase at higher dopings via Luttinger's theorem later in the paper. 

{\it Thermodynamics and Entanglement.} 

We investigate the specific heat and thermal entropy given by,
\begin{equation}
    C = \frac{\partial E}{\partial T}, \quad \text{and} \quad S = \log (\mathcal{Z}) + \frac{E}{T},
\end{equation}
where $\mathcal{Z}$ denotes the canonical partition function, and 
$E = \langle H \rangle$ 
the internal energy. Results for system sizes
$N=12, 16, 18$ are shown in Fig.~\ref{fig:thermodynamics}. To obtain the results on
system sizes $N=16, 18$ we employed thermal pure quantum (TPQ) states~\cite{TPQ1,TPQ2} as described in Refs.~\cite{Wietek2018,Wietek2019,Honecker2020} 
similar to the finite-temperature Lanczos method~\cite{Jaklic1994,Prelovsek2013} 
(see \cite{supp} for details).
For each set of random couplings we sampled $R=5$ TPQ states, cf. \cite{Wietek2019}.
Error estimates have been obtained from $1000, (400, 100)$ random couplings for
$N=12, (16, 18)$.

The specific heat for $p=0$ exhibits in Fig.~\ref{fig:thermodynamics}(a) exhibits 
a broad maximum at $T \approx 0.25$, in agreement with previous
results~\cite{MJR02}. At small values of doping $p \lesssim 1/6$ this maximum 
remains at $T \approx 0.25$ while we observe an increase of the specific heat
at higher temperatures. The maximum is gradually shifted towards a higher value
$T \approx 0.50$ for dopings from $p=1/4$ to $p=1/2$. At low temperatures we 
observe that the specific heat is approximately linear in temperature, with a maximal
slope attained between dopings $p=0.20$ and $p=0.40$. The linear-in-$T$ coefficient
of the specific heat, $\gamma= C/T$, is shown in Fig.~\ref{fig:thermodynamics}(b).
We observe an increase of $\gamma$ when lowering the temperature for all values
of doping. We show $\gamma$ at temperature $T = 0.05$ as a function of doping in Fig.~\ref{fig:thermodynamics}(c) for $N=12, 16, 18$. At this temperature, the 
maximum is attained at $p\approx 0.25$. However, we find that this maximum is 
dependent on the temperature. At temperatures below $T=0.05$ sample fluctuations
become too large for a reliable estimate of the maximum. 
We note that a divergence of the $\gamma$ coefficient has been reported 
at the pseudogap quantum critical point in cuprate superconductors ~\cite{Michon18}.

\begin{figure}[t]
    \centering
    \includegraphics[width=\columnwidth]{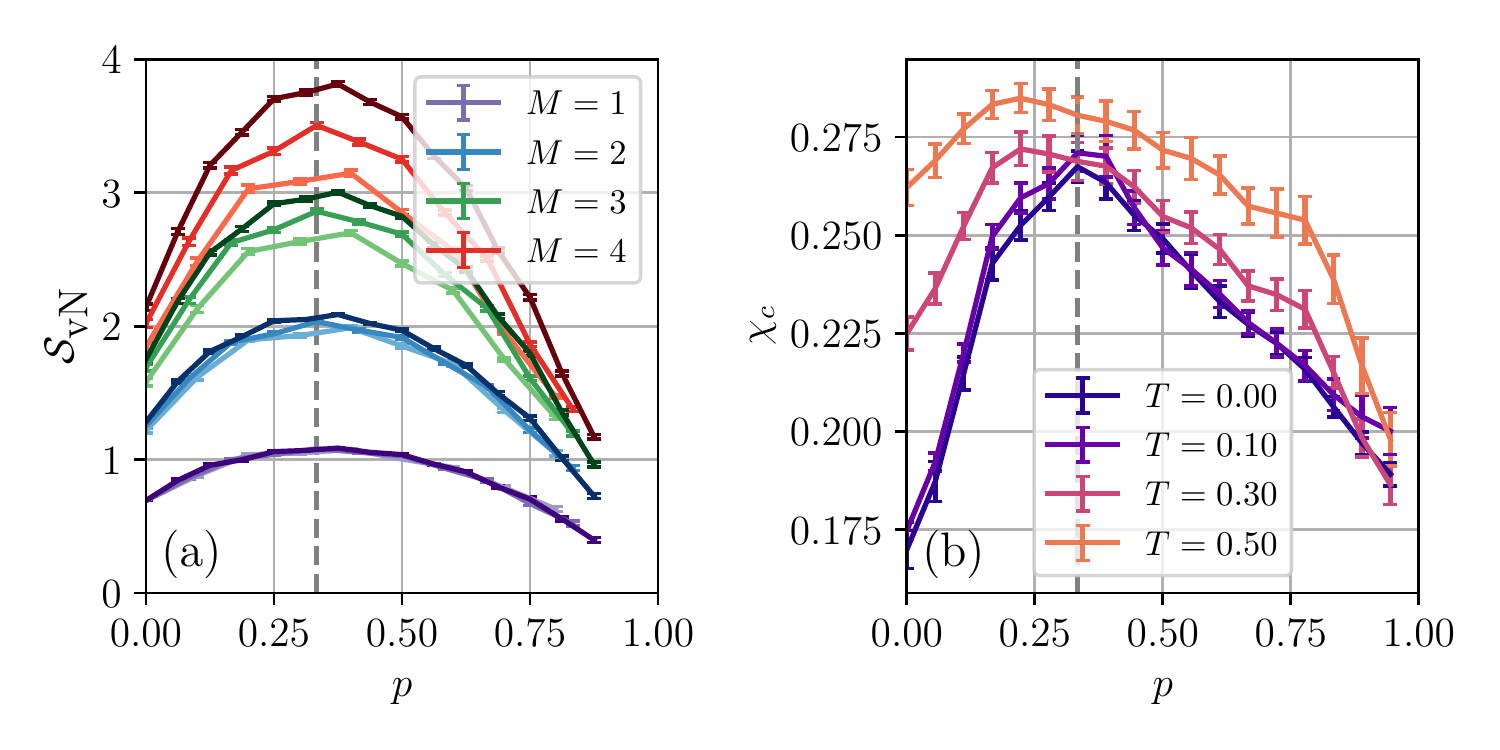}
    \caption{(a) The ground state entanglement entropy
    $\mathcal{S}_{\textrm{vN}}$ of subsystems of size $M$. Results are compared for total system size $N=10, 12, 16$,
    shown as increasing opacity. The maxima are attained at values close
    to $p = 1/3$, indicated by the gray dashed line. 
    (b) Charge susceptibility $\chi_c$ for different temperatures at $N=18$. The low-temperature maximum at doping $p=1/3$ is shifted towards a smaller doping
    $p\approx 0.2$ at higher temperatures.}
    \label{fig:entanglement}
\end{figure}

The thermal entropy for different temperatures and $N=12, 16, 18$ is shown in
Fig.~\ref{fig:thermodynamics}(d). Again we observe maxima at dopings between 
$p=0.20$ and $p=0.40$ depending on temperature. At $T=0.05$ the maximum is
attained at 
\begin{equation}
    \label{eq:pcthermal}
    \tilde{p} \approx 0.296 \pm 0.025.
\end{equation}
We refer to the supplement~\cite{supp} for more discussion of the $T$ dependence of the thermal entropy. To access the limit $T\rightarrow 0$ we calculate the von-Neumann entanglement entropy of the ground state,
\begin{equation}
    \mathcal{S}_{\textrm{vN}}(A) = -\Tr [ \rho_A \log \rho_A].
\end{equation}
Here, $\rho_A = \Tr_B(\ket{\psi_0}\bra{\psi_0})$ is the reduced density matrix of 
the ground state $\ket{\psi_0}$ on a subsystem $A$. $B$ denotes the
complement of $A$. Results for $\mathcal{S}_{\textrm{vN}}(A)$ for subsystem 
sizes $M=1,2,3,4$ and total system sizes $N=10,12, 16$ are shown in
Fig.~\ref{fig:entanglement}. We find that the single-site ($M=1$) and two-site
($M=2$) entanglement entropy are well converged as a function of total system size
$N$. 
For a $N=16$ site cluster and $M=4$ we estimate we estimate the maximum to be located at, 
\begin{equation}
    \label{eq:pcentanglement}
    \tilde{p} \approx 0.285 \pm 0.024 \quad [\text{from }\mathcal{S}_{\textrm{vN}}(A)],
\end{equation}
in agreement with our estimate obtained from the thermal entropy at $T=0.05$ in
Eq.~\ref{eq:pcthermal}.

Finally, we investigate the charge susceptibility (compressibility),
\begin{equation}
    \chi_c = \frac{\partial n}{ \partial \mu} = 
    \left( \frac{\partial^2 e}{\partial n^2} \right)^{-1} = 
    \left( \frac{\partial^2 e}{\partial p^2} \right)^{-1},
\end{equation}
computed by taking the inverse of the second derivative of the internal 
state energy density $e = E/N$ w.r.t. doping $p$. Here, the chemical potential 
is given by $\mu = \partial e / \partial n$. Results for different temperatures 
at $N=18$ are shown in Fig.~\ref{fig:entanglement}(b). At temperatures $T=0$ and
$T=0.1$ we detect a maximum at doping $p=1/3$. We observe a shoulder-like feature
at lower doping. At higher temperatures $T=0.3$ and $T=0.5$ this
feature develops into a maximum at $p\approx 0.2$. We notice, that this shift 
matches the shift of $\tilde{p}$ in the thermal entropy shown in
Fig.~\ref{fig:thermodynamics}(b,c).
We note that the occurrence of a maximum in the compressibility, specific heat coefficient and 
local entanglement entropy has been recently 
discussed in cluster-DMFT studies of the Hubbard model 
without randomness in relation to the pseudogap and Mott critical points~\cite{fratino2016,sordi2019,walsh2019,walsh2020}.

\begin{figure}[t]
    \centering
    \includegraphics[width=0.5\textwidth]{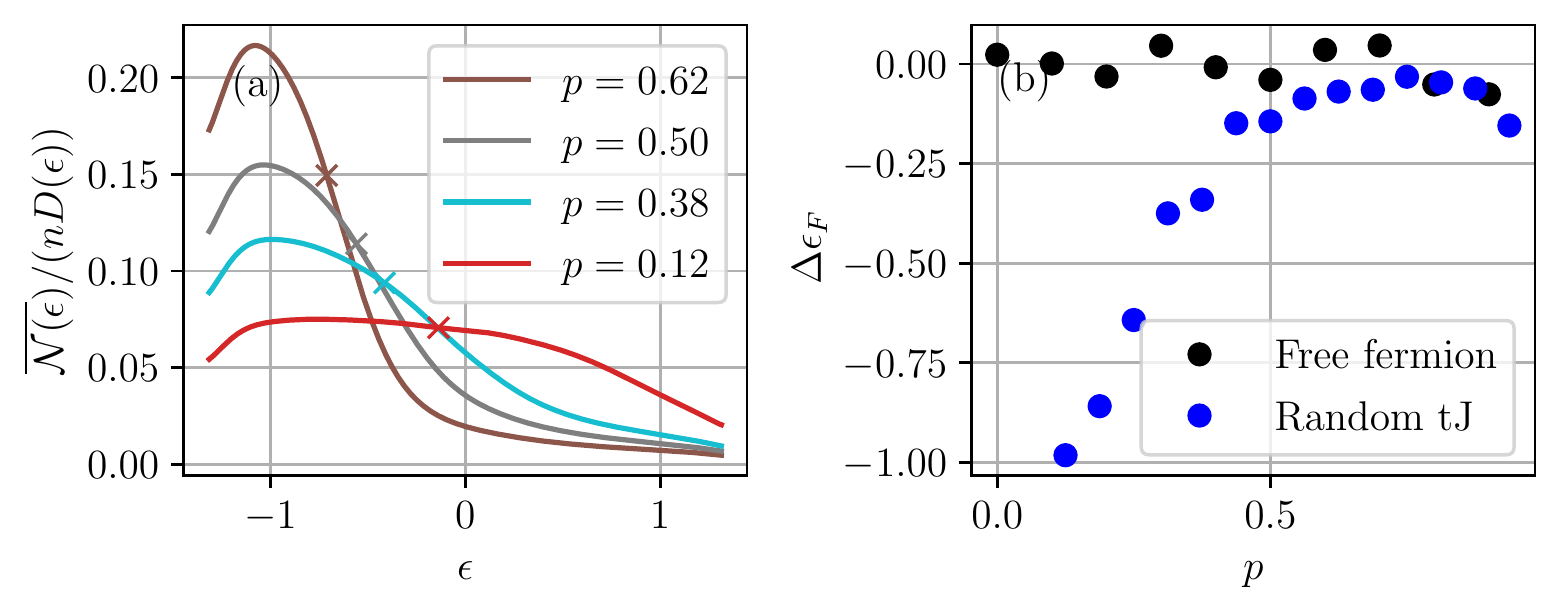}
    \caption{
    (a) At high values of doping, the one-particle energy distribution function drops sharply near the energy level predicted by Luttinger's theorem (marked by crosses). At lower values of doping, this function becomes more broadened, suggesting a breakdown of Luttinger's theorem. (b) A comparison of the Fermi energy given by Luttinger's theorem and the numerically-computed value given by the inflection point of the one-particle energy distribution function. For a 16 site cluster, the two show good agreement up to a critical value between $6/16 = 0.38$ and $7/16 = 0.44$, in contrast with the same quantity computed for free fermions which agree well for all values of doping.}
    \label{fig:luttingerRelation}
\end{figure}

{\it Luttinger's theorem.} Having found strong signatures of a spin glass phase persisting from half filling up to $p_c \approx 1/3$, we now provide evidence of a Fermi liquid phase at higher values of doping, which vanishes at a critical value of doping near the onset of spin glass order. To verify the presence of a Fermi liquid phase, we introduce the \textit{one-particle energy distribution function},
\begin{equation}
{\cal N}(\epsilon) = \frac{1}{N} \sum_{\lambda}\delta(\epsilon - \epsilon_\lambda) \sum_{i j \sigma}  \braket{\lambda}{i} \langle c_{i \sigma}^\dagger c_{j \sigma} \rangle \braket{j}{\lambda}
    \label{eq:oneParticleEnergyDistribution}
\end{equation}
where $\ket{\lambda}$ are the single-particle \textit{non-interacting} eigenstates with energy $\epsilon_\lambda$, obtained by diagonalizing the hopping matrix $t_{ij}$. This quantity is analogous to the particle occupation number in momentum space, $n(\mathbf{k})$, commonly used in 
systems with translational invariance. 
For a non-interacting system with fixed particle number $n$, the averaged quantity $\bar{{\cal N}(\epsilon)}$ converges to $D(\epsilon) \theta(\epsilon - \epsilon_F)$, 
where $D(\epsilon)$ is the single-particle density of states and $\epsilon_F$ is the Fermi energy, defined by:
\begin{equation}
    D(\epsilon)\,=\,\frac{1}{N}\overline{\sum_\lambda \delta(\epsilon-\epsilon_\lambda)}\,\,\,,\,\,\,
    n = 2 \int_{-\infty}^{\epsilon_F} \dd{\epsilon} D(\epsilon)\,.
    \label{eq:fermiEnergyDef}
\end{equation}
For the interacting system, we show in the supplemental material~\cite{supp} that, because the random couplings are all to all, ${\cal N}(\epsilon)$ displays self-averaging properties in the thermodynamic limit $N\rightarrow\infty$. 
In this limit, the signature of Luttinger's theorem is a discontinuity of $\bar{{\cal N}(\epsilon)}$ at the non-interacting 
value of $\epsilon_F$ defined in Eq.~(\ref{eq:fermiEnergyDef}).

In Fig.~\ref{fig:luttingerRelation}, we plot the quantity $\bar{{\cal N}(\epsilon)}/D(\epsilon)$, averaged over $1000$ realizations on a 16-site cluster. The density of states $D(\epsilon)$ is a semicircle distribution in the large-$N$ limit; however, in order to account for finite-size corrections to this distribution, we instead use the numerically calculated value of $D(\epsilon)$ obtained from our data. Although the drop in particle occupation at the Fermi energy is substantially broadened due to interactions and finite-size effects, the location of the inflection point still reliably tracks the location of the Fermi energy predicted by Luttinger's theorem at high values of doping as shown in Fig.~\ref{fig:luttingerRelation}. The effects of the infinite-strength Hubbard repulsion becomes stronger at lower values of doping, eventually causing a breakdown of Luttinger's theorem at a critical doping $0.38 < p_c < 0.44$, which is also the location where spin glass order appears to emerge. 

{\it Discussion and Conclusion.}
Our numerical results demonstrate a transition in the random all-to-all $t$-$J$ model from a spin glass 
to a disordered Fermi liquid at a critical value of doping. 
The near-critical behavior has similarities to the criticality of SYK models, consistent with recent theoretical proposals~\cite{randomtJ1} and numerical results on related systems~\cite{Cha19}. 
We find a near-critical dynamic spin susceptibility which is consistent with the SYK behavior $\chi''(\omega) \sim \mbox{sgn}(\omega) \left[1 - g |\omega| + \ldots \right]$ over a significant frequency regime; the $g$ term is a universal boundary ``graviton'' contribution. This is the first appearance of such features in a doped spin-1/2 SU(2) model. 
SYK criticality also predicts an extensive zero temperature entropy: we do find a maximum in the entropy near the critical point, but our finite-size data does not allow us to identify if there is an extensive contribution. However, we note that for SU$(M=2)$ models, SYK criticality is pre-empted at small enough $T$ by a spin glass instability \cite{GPS01,dumi2020}, and so the extensive $T=0$ entropy is not ultimately expected.
We also find a maximum in the entanglement entropy, specific heat coefficient, and compressibility near criticality.

An interesting observation is that the breakdown of Luttinger's theorem coming from high doping, as well as the vanishing of spin glass order from low doping, occurs near $p=0.4$, which differs from the maxima in the thermodynamic and entanglement entropy closer to $p=0.3$. While the system sizes accessible to our methods are relatively small and only discrete values of doping are accessible, recent (E)DMFT calculations of the $t$-$J$ model with finite Hubbard repulsion~\cite{dumi2020} also give evidence of SYK criticality occurring at a lower value of doping than the spin glass/Fermi liquid transition. 
These observations are consistent with the spin glass instability of SYK criticality for finite $M$ \cite{GPS01} noted above.
Understanding the nature of this separation, and the very low $T$ at which the spin glass instability of SYK criticality appears, remain open questions to be explored.

{\it Acknowledgements.} We thank P.~Dumitrescu, O.~Parcollet, M.~Rozenberg and N.~Wentzell for valuable discussions.
This research was supported by the National Science Foundation under Grant No. DMR-2002850. 
AG acknowledges the support of the European Research Council (ERC-319286-QMAC). This work was also supported by the Simons Collaboration on Ultra-Quantum Matter, which is a grant from the Simons Foundation (651440, S.S.).
The Flatiron Institute is a division of the Simons Foundation.\\
H.S. and A.W. contributed equally to this work.
\bibliography{main}



\section*{Supplementary material}

# Supplement to
# Quantum phase transition at non-zero doping in a random $t$-$J$ model

Henry Shackleton,[1] Alexander Wietek,[2] Antoine Georges,[2,3,4,5] and Subir Sachdev[1]

[1]*Department of Physics, Harvard University, Cambridge MA 02138, USA*
[2]*Center for Computational Quantum Physics, Flatiron Institute, New York, NY 10010 USA*
[3]*Collège de France, 11 place Marcelin Berthelot, 75005 Paris, France*
[4]*CPHT, CNRS, École Polytechnique, IP Paris, F-91128 Palaiseau, France*
[5]*DQMP, Université de Genève, 24 quai Ernest Ansermet, CH-1211 Genève, Suisse*

## SPIN GLASS ANALYSIS

As described in the main text, the spectral function of the random $t$-$J$ model near half filling has a peak at low frequency, suggesting spin glass order. To establish this rigorously, one must show that the variance of the peak goes to zero in the thermodynamic limit while the integrated spectral weight remains non-zero, indicating delta function-like behavior. We isolate the low-frequency peak by subtracting off a background contribution, given by the large-$M$ solution of the SY model. We then fit the remaining low-frequency peak to the function

$$\chi''_{low}(\omega) = \omega C \exp\left[-\frac{\omega^2}{2\sigma^2}\right] . \qquad (1)$$

In Fig. 1, we show the extrapolation of $\Gamma$ to the thermodynamic limit for several values of doping up to $p = 1/3$. As expected of spin glass behavior, $\Gamma$ vanishes in the thermodynamic limit. This is in contrast with the integrated spectral weight of Eq. 1, which we show in the main text is non-zero in the the thermodynamic limit and corresponds to the Edwards-Anderson spin glass order parameter $q$.

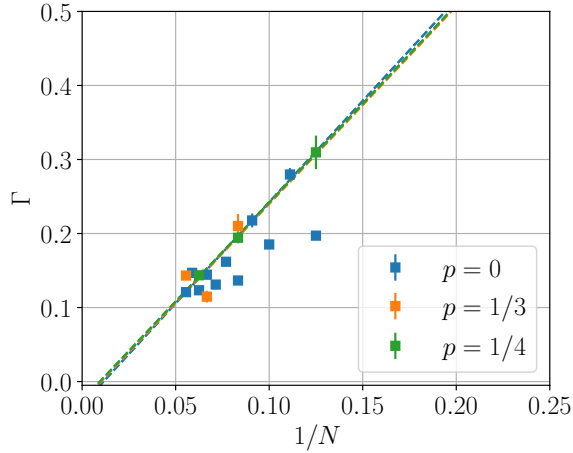

FIG. 1. At low dopings, the low-frequency peak in the spectral function can be isolated and fit to Eq. 1. In the thermodynamic limit, we confirm that the variance $\Gamma$ vanishes up to $p = 1/3$. Due to a prominent even/odd particle effect at half filling, we only extrapolate $\Gamma$ at half-filling for even system sizes.

## THERMAL PURE QUANTUM STATES

The computation of thermodynamic quantities in the main text has been performed using thermal pure quantum states [1, 2] together with the Lanczos algorithm. This allowed us to reach system sizes beyond the reach of full exact diagonalization. This approach is closely related to the finite-temperature Lanczos method [3, 4]. We will now briefly explain the method. The trace of any operator $H$ can be evaluated by taking random average values,

$$\text{Tr}(A) = D \overline{\langle r \,|\, A \,|\, r \rangle}, \qquad (2)$$

where $|r\rangle$ is a normalized random vector, $\langle r\,|\,r\rangle = 1$, with normal distributed coefficients, $\langle m\,|\,r\rangle \sim \mathcal{N}(0,1)$, and $D$ denotes the dimension of the Hilbert space. Here, $\{|m\rangle\}_{m=1,\ldots D}$ denotes an arbitrary orthonormal basis of the Hilbert space and $\overline{\cdots}$ denotes averaging over random realizations of $|r\rangle$. Hence, a thermal expectation value of an observable $\mathcal{O}$ can be written as as,

$$\langle \mathcal{O} \rangle = \frac{1}{\mathcal{Z}} \text{Tr}(\mathrm{e}^{-\beta H} \mathcal{O}) = \frac{\overline{\langle \beta\,|\,\mathcal{O}\,|\,\beta\rangle}}{\overline{\langle \beta\,|\,\beta\rangle}}, \qquad (3)$$

where $\mathcal{Z} = \text{Tr}(\mathrm{e}^{-\beta H})$ denotes the partition function and we define the so-called *thermal pure quantum* (TPQ) state [1, 2] at inverse temperature $\beta = 1/T$,

$$|\beta\rangle = \mathrm{e}^{-\beta H/2} |r\rangle. \qquad (4)$$

This way, thermal expectation values can be evaluated efficiently using the Lanczos algorithm, whereas the exact computation of the trace of an exponential requires full diagonalization. In the main text we present data for the specific heat, internal energy and entropy, which are all computed from expectation values of powers of the Hamiltonian with TPQ states of the form $\langle \beta\,|\,H^k\,|\,\beta \rangle$. Using the Lanczos algorithm this quantity is efficiently approximated by,

$$\langle \beta\,|\,H^k\,|\,\beta\rangle \approx e_1^\dagger \mathrm{e}^{-\frac{\beta}{2} T_n} T_n^k \mathrm{e}^{-\frac{\beta}{2} T_n} e_1, \qquad (5)$$

where $e_1 = (1, 0, \ldots, 0)^\dagger$ and $T_n$ denotes the tridiagonal matrix of the Lanczos algorithm after $n$ steps. The convergence is typically exponentially fast, such that results can be attained up to machine precision. For a more detailed description of the method we refer the reader to Ref. [5]. We notice in Eq. (5), that once the Lanczos algorithm has been applied to compute the tridiagonal matrix, results can be derived at all temperatures simultaneously without rerunning the expensive Lanczos algorithm.

Instead of one single computation as done for evaluating a trace, using TPQ states requires us to to perform random sampling with multiple vectors $|r\rangle$ and compute error estimates. Since expectation values of the form Eq. (3) are non-linear in $|r\rangle$, we perform jackknife resampling [6] of the data. Interestingly, larger system sizes typically require less random realizations $|r\rangle$ to obtain comparable errorbars. Refs. [1, 2] give a mathematical proof, that for a constant free energy density, the variance of the estimate in Eq. (3) is exponentially small in the system size. In the main text we typically average over $R = 5$ random realizations of the TPQ states.

## TEMPERATURE DEPENDENCE OF THE THERMAL ENTROPY

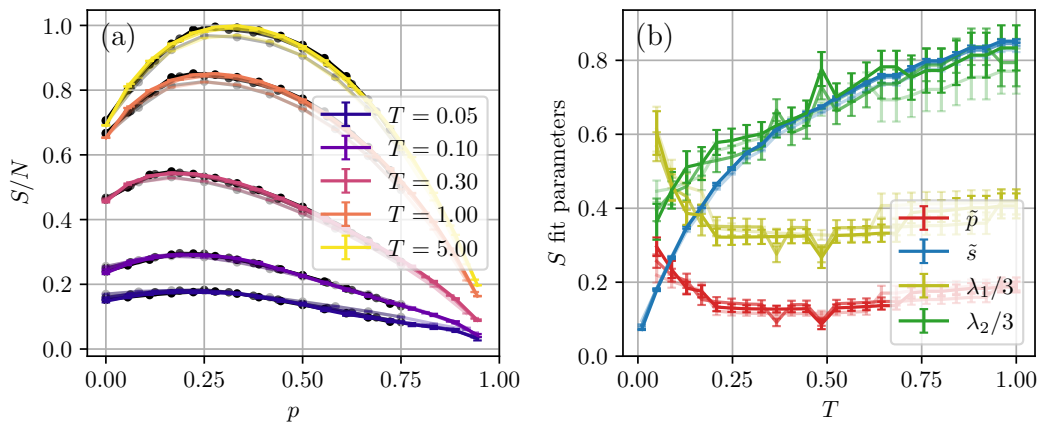

FIG. 2. (a) The thermal entropy $S$ as a function of doping for various temperatures. Black dots show the ansatz Eq. 6 at optimal fitting parameters. (b) Estimates of the parameters in Eq. 6. $\tilde{p}$ corresponds to the doping value with maximal entropy, $\tilde{s}$ corresponds to the maximal entropy density.

In the limit $T \to \infty$ the thermal entropy $S$ attains a maximum exactly at $p = 1/3$ for $N \to \infty$ in the canonical

ensemble. The ansatz,

$$S/N = \begin{cases} -K\,|p-\tilde{p}|^{\lambda_1} + \tilde{s} & \text{for } p \leq \tilde{p} \\ -K\,|p-\tilde{p}|^{\lambda_2} + \tilde{s} & \text{for } p > \tilde{p} \end{cases}, \tag{6}$$

is found to describe our entropy data with considerable precision. A comparison between the ansatz (black circles) and the ED data is shown in Fig. 2(a). The parameters $\tilde{p}$, $\tilde{s}$, $\lambda_1$, $\lambda_2$, and $K$ are fitted for dopings $p \in [0, 0.75]$ using the Levenberg-Marquardt algorithm [7, 8], from which we obtain an (error) estimate of the parameters, shown in Fig. 2(e).

Our estimate of $\tilde{p}$ is increasing when lowering the temperature below $T = 0.25$. At $T = 0.05$ and $N = 18$ we obtain an estimate,

$$\tilde{p} \approx 0.296 \pm 0.025 \quad [\text{from } S(T = 0.05)]. \tag{7}$$

This value is consistent with the maximum of $\gamma$, observed in the main text. However, we find that both increasing the system size and lowering temperature increases our estimate of the critical doping $\tilde{p}$ when estimated as above. At temperatures below $T = 0.05$ estimates are found to be unreliable due to sample fluctuations.

## SELF-AVERAGING, ELECTRON DISTRIBUTIONS, AND THE LUTTINGER THEOREM

### Self-averaging from the cavity method

In this section, we establish that, in the thermodynamic limit $N \to \infty$, some local observables have *self-averaging* properties in this fully connected random model. This means that, when considered for a given site, they converge with probability one to their average over samples. We also establish the connection to extended dynamical mean-field equations (EDMFT) that allow for a direct study of the model in the thermodynamic limit. We *do not* consider the spin-glass phase in this section.

For the sake of generality, we consider the finite-$U$ version of the model, the $t$-$J$ limit corresponding to $U = \infty$. The model is defined on a fully connected lattice of $N$ sites by the Hamiltonian:

$$H = -\sum_{ij,\sigma=\uparrow,\downarrow} t_{ij}\,c_{i\sigma}^\dagger c_{j\sigma} + U\sum_i n_{i\uparrow}n_{i\downarrow} - \sum_{i<j} J_{ij}\,\vec{S}_i \cdot \vec{S}_j \tag{8}$$

with:

$$t_{ij} = \frac{t}{\sqrt{N}}\,\varepsilon_{ij} \;,\; J_{ij} = \frac{J}{\sqrt{N}}\,\eta_{ij} \tag{9}$$

In these expressions, $\varepsilon$ and $\eta$ are random variables of zero mean and unit variance. The precise distribution is not important in the infinite-size (large-$N$) limit, as shown below.

Let us consider a fixed sample $\{\varepsilon_{ij}, \eta_{ij}\}$, and envision formally integrating over all sites except a single one (denoted by $i = 1$). The lattice with the 'cavity' removed (consisting of that site and all connections through $t_{1j}$ and $J_{1j}$) is in this case just a fully connected lattice of $N - 1$ sites. We follow the procedure in Ref. [9],Sec.III.A: in the large-$N$ limit, the effective action for site 1 obtained after integrating out all other d.o.fs is:

$$S_{\text{eff}}[1] = -\int_0^\beta d\tau d\tau' \sum_\sigma c_{1\sigma}^\dagger(\tau)\left(\delta(\tau-\tau')(-\partial_\tau + \mu) - \Delta_1(\tau-\tau')\right)c_{1\sigma}(\tau') + \\ + U\int_0^\beta d\tau\,n_{1\uparrow}n_{1\downarrow} - \tfrac{1}{2}\int_0^\beta d\tau d\tau'\,Q_1(\tau-\tau')\vec{S}_1(\tau)\cdot\vec{S}_1(\tau'). \tag{10}$$

All higher order correlators are lower order in $1/N$ (finite-size corrections). The dynamical mean-fields, i.e the hybridisation function $\Delta_1$ and retarded spin-spin interaction $Q_1$ are given by the cavity equations:

$$\Delta_1 = \frac{t^2}{N}\sum_{i,j\neq 1}\varepsilon_{1i}\varepsilon_{1j}\,G_{ij}^{[1]} \;,\; Q_1 = \frac{J^2}{N}\sum_{i,j\neq 1}\eta_{1i}\eta_{1j}\,\chi_{ij}^{[1]} \tag{11}$$

In this expression $G_{ij}(\tau - \tau') \equiv -\langle T c_i^\dagger(\tau) c_j(\tau')\rangle$ and $\chi_{ij} \equiv \langle \vec{S}_i \cdot \vec{S}_j\rangle/3$ are the Green's function and spin-spin correlation function. The superscript $G^{[1]}$ means that we are considering these quantities for the subsystem of $N-1$ spins remaining once the cavity (site 1 and its connections) has been created.

Let us analyze Eqs. (11) for the hybridisation function, separating diagonal and off-diagonal terms:

$$\Delta_1 = \frac{t^2}{N} \sum_{i=2}^{N} \varepsilon_{1i}^2 \, G_{ii}^{[1]} + \frac{t^2}{N} \sum_{i,j \neq 1; i \neq j} \varepsilon_{1i} \varepsilon_{1j} \, G_{ij}^{[1]} \tag{12}$$

The key point is that $G_{ij}^{[1]}$ *do not depend on the random variables* $\varepsilon_{1i}$. Taking the $N \to \infty$ limit amounts to take a disorder average of these terms, and because of this independence, the average applies separately to $\varepsilon_{1i}$ and $G_{ii}^{[1]}$. The second term ($i \neq j$) averages out to zero, and the first one yields finally:

$$\Delta_1 = t^2 \lim_{N \to \infty} \sum_i G_{ii} = t^2 \, \overline{G} \, , \quad (N \to \infty) \tag{13}$$

In which the overline denotes an average over samples. We are assuming here that there is no 'ergodicity breaking' in the phase being considered: the average over sites is equivalent to an average over samples. Hence, the dynamical mean field $\Delta$ does not depend on the specific site or on the specific sample, in the infinite size limit: it *self-averages*. A similar reasoning applies to $Q$. Finally the self-consistent equations read for infinite size (I am dropping the overlines and site index everywhere when there is no possible confusion):

$$\Delta(i\omega_n) = t^2 \, G(i\omega_n) \, , \quad Q(i\omega_n) = J^2 \, \chi(i\omega_n) \tag{14}$$

in which the local ($i = j$) correlators $G$ and $\chi$ have to be calculated with the effective action (10) - that's the EDMFT construction.

For completeness, we recall that the (local) self-energy in the infinite-volume limit is given by the difference between the inverse of the interacting and non-interacting Green's functions, namely:

$$\Sigma(i\omega_n) = i\omega_n + \mu - \Delta(i\omega_n) - G^{-1}(i\omega_n) = i\omega_n + \mu - t^2 G - G^{-1} \tag{15}$$

$\Delta$ and the local $G_{ii}$ being self-averaging, $\Sigma_{ii}$ also is.

Incidentally, for the non-interacting system ($U = J = 0$), $\Sigma = 0$ and the solution of the quadratic equation: $z - t^2 G - G^{-1} = 0$ yields the non-interacting local Green's function ($z \equiv i\omega_n + \mu$, but the formula is valid for any $z$ in the complex plane):

$$G_0(z) = \frac{1}{2t^2} \left[ z - \text{sign}[\text{Im}(z)] \sqrt{z^2 - 4t^2} \right] = \int d\varepsilon \, \frac{D_\infty(\varepsilon)}{z - \varepsilon} \tag{16}$$

from which the (Wigner) semi-circular distribution immediately follows:

$$D_\infty(\varepsilon) \equiv -\frac{1}{\pi} \text{Im} G_0(\varepsilon + i0^+) = \frac{1}{2\pi t^2} \sqrt{4t^2 - \varepsilon^2} \, , \quad \varepsilon \in [-2t, +2t] \tag{17}$$

### Green's function and one-particle energy distribution

We can use the eigenstates of the one-particle non-interacting problem ($U = J_{ij} = 0$) as a *basis set* to represent any single-particle correlation function of the interacting problem. These states are defined by, for a given sample $t_{ij}$:

$$\hat{t}|\lambda\rangle = \varepsilon_\lambda |\lambda\rangle \, , \quad \text{i.e.} \sum_j t_{ij} \langle j | \lambda \rangle = \varepsilon_\lambda \langle i | \lambda \rangle \tag{18}$$

The Fock space of the many-body problem is constructed as the number occupancy states $|\{n_\lambda\}\rangle$ and is a full basis for the many-body problem. The single-particle DOS of the non-interacting system reads:

$$D_{ij}(\varepsilon) = \frac{1}{N} \sum_\lambda \delta(\varepsilon - \varepsilon_\lambda) \, \langle i | \lambda \rangle \langle \lambda | j \rangle \tag{19}$$

In the $N \to \infty$ limit, $D_{ii}$ converges (and self-averages) to the semi-circular DOS $D_\infty$ defined above.

Consider now the interacting system, for a given sample $t_{ij}, J_{ij}$ and finite $N$. We define the one-electron Green's function in the usual way $G_{ij}(\tau - \tau') \equiv -\langle T c_i^\dagger(\tau) c_j(\tau')\rangle$, but it can actually be viewed as a one-body operator $\hat{G}$ that we can look at in any basis set, for example in the eigenstate basis:

$$G_{\lambda\lambda'}(i\omega_n) = \left\langle \lambda \left| \hat{G} \right| \lambda' \right\rangle = \sum_{ij} \langle \lambda | i \rangle\, G_{ij}(i\omega_n)\, \langle j | \lambda' \rangle \qquad (20)$$

Note that for a given sample and finite $N$, this is not diagonal in $\lambda$. Correspondingly, a self-energy $\sigma_{ij}$ can be defined as (we are careful to use a different notation here, since this is for a given sample and finite $N$):

$$\hat{G}^{-1} \equiv i\omega_n + \mu - \hat{t} - \hat{\sigma} \;, \text{in site basis}: \; [G^{-1}]_{ij} = (i\omega_n + \mu)\delta_{ij} - t_{ij} - \sigma_{ij} \qquad (21)$$

Things get simpler in the infinite-volume limit. The off-diagonal components of the self-energy $\sigma_{i\neq j}$ vanish, and the diagonal ones self-average and converge to the local self-energy defined above: $\sigma_{ii} \to \Sigma$. Hence, the expression of the Green's function becomes:

$$[G^{-1}]_{ij} = [i\omega_n + \mu - \Sigma(i\omega_n)]\,\delta_{ij} - t_{ij} \;, \; (N \to \infty) \qquad (22)$$

Note that off-diagonal components of the Green's functions *do not self-average*. They are individually of typical order $1/\sqrt{N}$, but we have to take them into account when calculating the kinetic energy for example, since we sum over all bonds. Given (22), the Green's function for $N = \infty$ now acquires a simple *diagonal* representation in the basis of eigenstates of $\hat{t}$:

$$G_{ij}(i\omega_n) = \sum_\lambda \langle i | \lambda \rangle \frac{1}{i\omega_n + \mu - \Sigma(i\omega_n) - \varepsilon_\lambda} \langle \lambda | j \rangle \qquad (23)$$

It is convenient to define the (sample independent) Green's function for a given energy $\varepsilon$ in the semi-circular 'band' as:

$$G(i\omega_n, \varepsilon) \equiv \frac{1}{i\omega_n + \mu - \Sigma(i\omega_n) - \varepsilon} \qquad (24)$$

which is the natural quantity we would routinely look at in the EDMFT framework. The connection between this and the Green's function for a given sample, in the infinite size limit $N = \infty$, is given by:

$$G_{ij}(i\omega_n) = \int d\varepsilon\, D_{ij}(\varepsilon)\, G(i\omega_n, \varepsilon) \;, \; (N = \infty) \qquad (25)$$

Let us now consider (for a given sample and any $N$) the one-body distribution function:

$$N_\lambda = \langle \hat{n}_\lambda \rangle = \sum_{ij} \langle \lambda | i \rangle \sum_\sigma \left\langle c_{i\sigma}^\dagger c_{j\sigma} \right\rangle \langle j | \lambda \rangle \qquad (26)$$

In the non-interacting case, the ground-state is a Slater determinant of the $\lambda$ states, and hence at $T = 0$ $N_\lambda = 1$ for all filled states and 0 for empty states. We can more conveniently look at it by filtering in energy and define:

$$\mathcal{N}(\varepsilon) = \frac{1}{N} \sum_\lambda \delta(\varepsilon - \varepsilon_\lambda) N_\lambda = \frac{1}{N} \sum_\lambda \delta(\varepsilon - \varepsilon_\lambda) \sum_{ij\sigma} \langle \lambda | i \rangle \left\langle c_{i\sigma}^\dagger c_{j\sigma} \right\rangle \langle j | \lambda \rangle \qquad (27)$$

which can also be written:

$$\mathcal{N}(\varepsilon) = \sum_{ij\sigma} D_{ij}(\varepsilon) \left\langle c_{i\sigma}^\dagger c_{j\sigma} \right\rangle \qquad (28)$$

obeying (with $n$ the electron density):

$$\int d\varepsilon\, \mathcal{N}(\varepsilon) = \frac{N_e}{N} = n \qquad (29)$$

We can also sample average and consider $\overline{\mathcal{N}(\varepsilon)}$.

Now we establish the connection, in the $N = \infty$ limit, between this distribution function and what we would naturally calculate in the EDMFT context, which is the number distribution function as a function of the single-particle energy:

$$N(\varepsilon) = 2G(\tau = 0^-, \varepsilon) = 2\frac{1}{\beta}\sum_n e^{i\omega_n 0^+}\frac{1}{i\omega_n + \mu - \Sigma(i\omega_n) - \varepsilon} \tag{30}$$

The factor of 2 is for the sum over spin. We note that:

$$\int d\varepsilon\, D_\infty(\varepsilon) N(\varepsilon) = n \tag{31}$$

Using (25), we obtain:

$$\overline{\mathcal{N}(\varepsilon)} = \frac{1}{N}\sum_\lambda \delta(\varepsilon - \varepsilon_\lambda) N(\varepsilon_\lambda) = D_\infty(\varepsilon) N(\varepsilon) \ , \ (N = \infty) \tag{32}$$

### Luttinger's theorem.

In the (self-averaging) infinite volume limit, the Green's function $G(\omega, \varepsilon)$ has a pole $\omega(\varepsilon)$ given by the quasiparticle equation:

$$\omega + \mu - \mathrm{Re}\Sigma(\omega + i0^+) = \varepsilon \tag{33}$$

where we have assumed that at low $\omega, T$ the imaginary part $\mathrm{Im}\Sigma(\omega + i0^+)$ is negligible. Hence the 'Fermi surface' (for a typical large sample) is located at:

$$\varepsilon_F = \mu - \mathrm{Re}\Sigma(0) \tag{34}$$

In the non-interacting system, the Fermi level $\varepsilon_F$ is given by:

$$n = 2\int_{-\infty}^{\varepsilon_F} d\varepsilon\, D(\varepsilon) \tag{35}$$

Hence, for a fixed density, the Luttinger theorem (Fermi 'surface' unchanged by interactions) translates into the following requirement:

$$\mu(n) - \mathrm{Re}\Sigma(0) = \varepsilon_F(n) \tag{36}$$

In the Fermi liquid phase, this can be established following the usual proof based on the existence of a Luttinger-Ward functional.

---



\end{document}